\documentclass[a4paper]{jpconf}
\usepackage{graphicx}
\begin{document}
\title{Multi-strange baryon production in pp, \mbox{p--Pb} and \mbox{Pb--Pb} collisions measured with ALICE}

\author{Domenico Colella for the ALICE Collaboration}

\address{Dipartimento Interateneo di Fisica ''M. Merlin'' and Sezione INFN, Via Orabona 4, 70126 Bari, Italy}

\ead{domenico.colella@cern.ch}

\begin{abstract}
Transverse momentum spectra and yields of charged $\Xi$ and $\Omega$ at mid-rapidity in pp, \mbox{p--Pb} and \mbox{Pb--Pb} collisions at the LHC have been measured by the ALICE Collaboration. These baryons are identified by reconstruction of their weak decay topology, in modes with only charged decay products, using the excellent tracking and particle identification capabilities of the detector. The recent measurements of the multi-strange baryon production relative to non-strange particles in \mbox{p--Pb} collisions are presented: this would help to understand the change in relative strangeness production from pp collisions to \mbox{Pb--Pb} collisions. Results on the nuclear modification factors for the charged $\Xi$ and $\Omega$ particles, compared with those for other light particles, are also reported.
\end{abstract}

\section{Introduction}
\label{sec-0}
Measuring strange and multi-strange particle production in relativistic heavy-ion interactions give a unique insight into the properties of the hot and dense matter created in the collision, as there is no net strangeness content in the initially colliding nuclei. Moreover, the high energy and luminosity available at the LHC grant a relative abundant production of such particles. 

An enhanced production of strange particles in \mbox{A--A} compared to pp interactions was one of the earliest proposed signatures of the formation of a deconfined Quark-Gluon Plasma (QGP) \cite{Rafelski}. In \mbox{pp} collisions, the production of strange particles is limited by the local canonical strangeness conservation. This is not the case in central heavy-ion collisions, where strange quarks are expected to be produced more abundantly, resulting in an enhanced production of strange baryons at the end of the  hadronisation stage. Strangeness enhancement in heavy-ion collisions with respect to \mbox{pp} (\mbox{p--Be}) has been observed in \mbox{Pb--Pb} collisions at the NA57 experiment \cite{strangEnhancNA57}, at STAR \cite{strangEnhancSTAR} in \mbox{Au--Au} collisions, and indeed confirmed by ALICE in central \mbox{Pb--Pb} collisions at \mbox{$\sqrt{s_{\rm NN}}$ = 2.76 TeV} \cite{strangEnhancALICE}.

One of the most striking differences between the \mbox{A--A} and \mbox{pp} collisions first observed at RHIC was the suppression of the hadron production at high transverse momentum ($p_{\rm T}$) in central \mbox{Au--Au} collisions at $\sqrt{s_{\rm NN}}$ = \mbox{200 GeV} when compared to expectations from an incoherent superposition of nucleon-nucleon interactions \cite{RHICRaa}. The effect, confirmed by the charged-particle nuclear modification factor ($R_{\rm AA}$) as a function of $p_{\rm T}$ measured in \mbox{Pb--Pb} collisions at the LHC \cite{ALICERaa}, is interpreted as the result of the energy loss suffered by partons when traversing the hot and dense matter created in an ultra-relativistic heavy-ion collision. The larger suppression observed at LHC compared to that at RHIC can be explained by the higher density of the medium created in collisions at higher centre-of-mass energy.

\section{Multi-strange decay reconstruction with the ALICE detector}
\label{sec-1}
The ALICE detector was designed to study heavy-ion physics at the LHC. It consists of a central barrel with a large solenoid providing a \mbox{0.5 T} field for tracking and particle identification and forward detectors for triggering and centrality selection. Tracking and vertexing are performed using the Inner Tracking System (ITS), consisting of six layers of silicon detectors, and the Time Projection Chamber (TPC). The two innermost layers of the ITS and the V0 detector (scintillation hodoscopes placed on either side of the interaction region) are used for triggering. The V0 also provides the centrality determination in \mbox{Pb--Pb} collisions as well as the multiplicity class selection in \mbox{p--Pb} collisions. A complete description of the ALICE sub-detectors can be found in \cite{JINST}.

Multi-strange baryons are reconstructed through their weak decay topologies, namely \mbox{$\Xi^{-} \rightarrow \pi^{-} + \Lambda$}, \mbox{$\Omega^{-} \rightarrow \textrm{K}^{-} + \Lambda$} (with \mbox{$\Lambda \rightarrow\pi^{-}  + \textrm{p}$}), and the corresponding charge conjugate decays for the anti-particles. The branching ratios are 63.9\% and 43.3\%, for the $\Xi$ and the $\Omega$, respectively. The $\Xi$ and the $\Omega$ candidates are found by combining reconstructed charged tracks: cuts on geometry and kinematics are applied to first select the $\Lambda$ candidate and then to match it with all the remaining secondary tracks (bachelor candidates). In addition, particle identification is performed by a selection on the specific energy loss in the TPC for the three daughter tracks.
 
\section{Results}
\label{sec-2}
\begin{figure}
\centering
\includegraphics[width=20pc,clip]{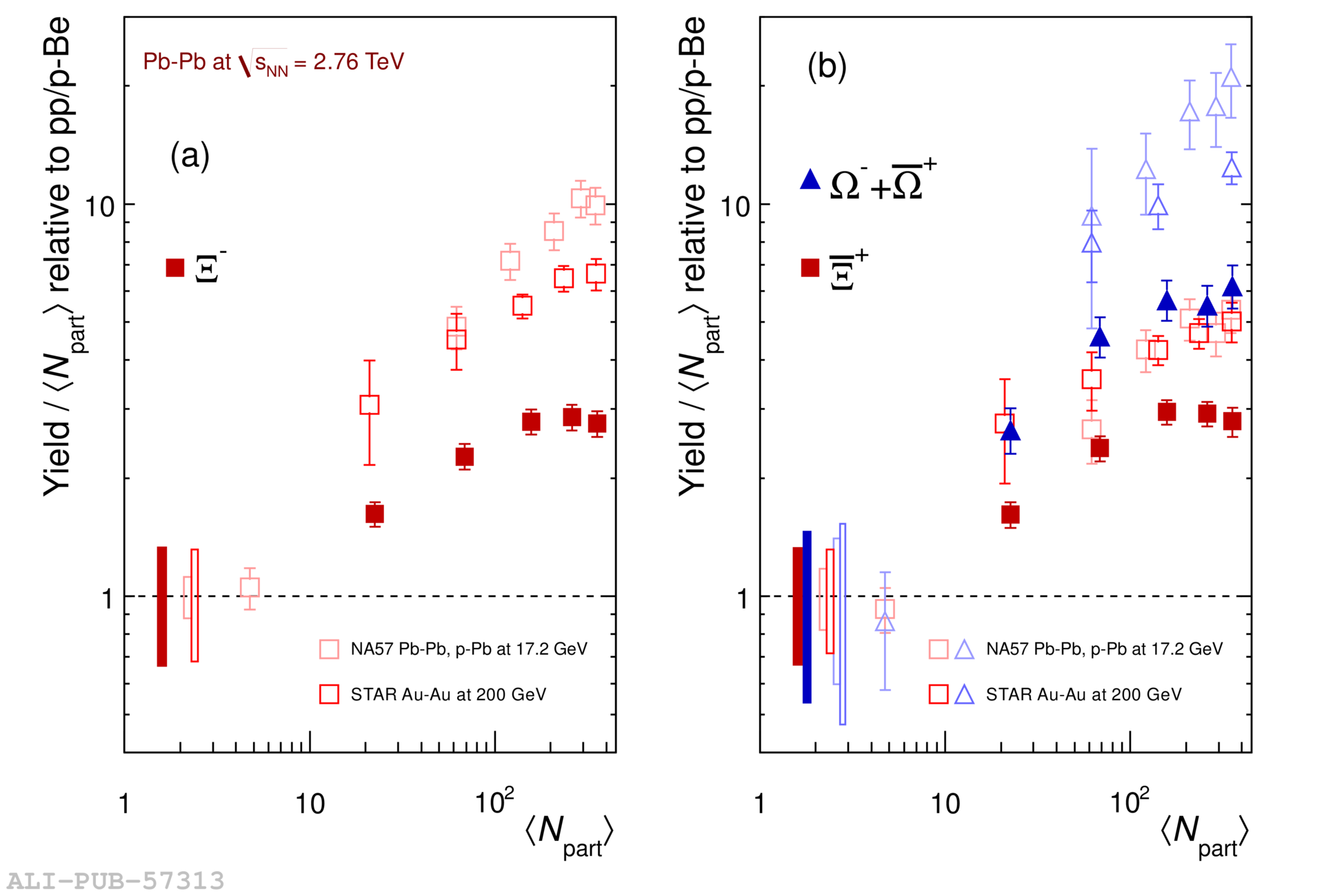}
\includegraphics[width=10pc,clip]{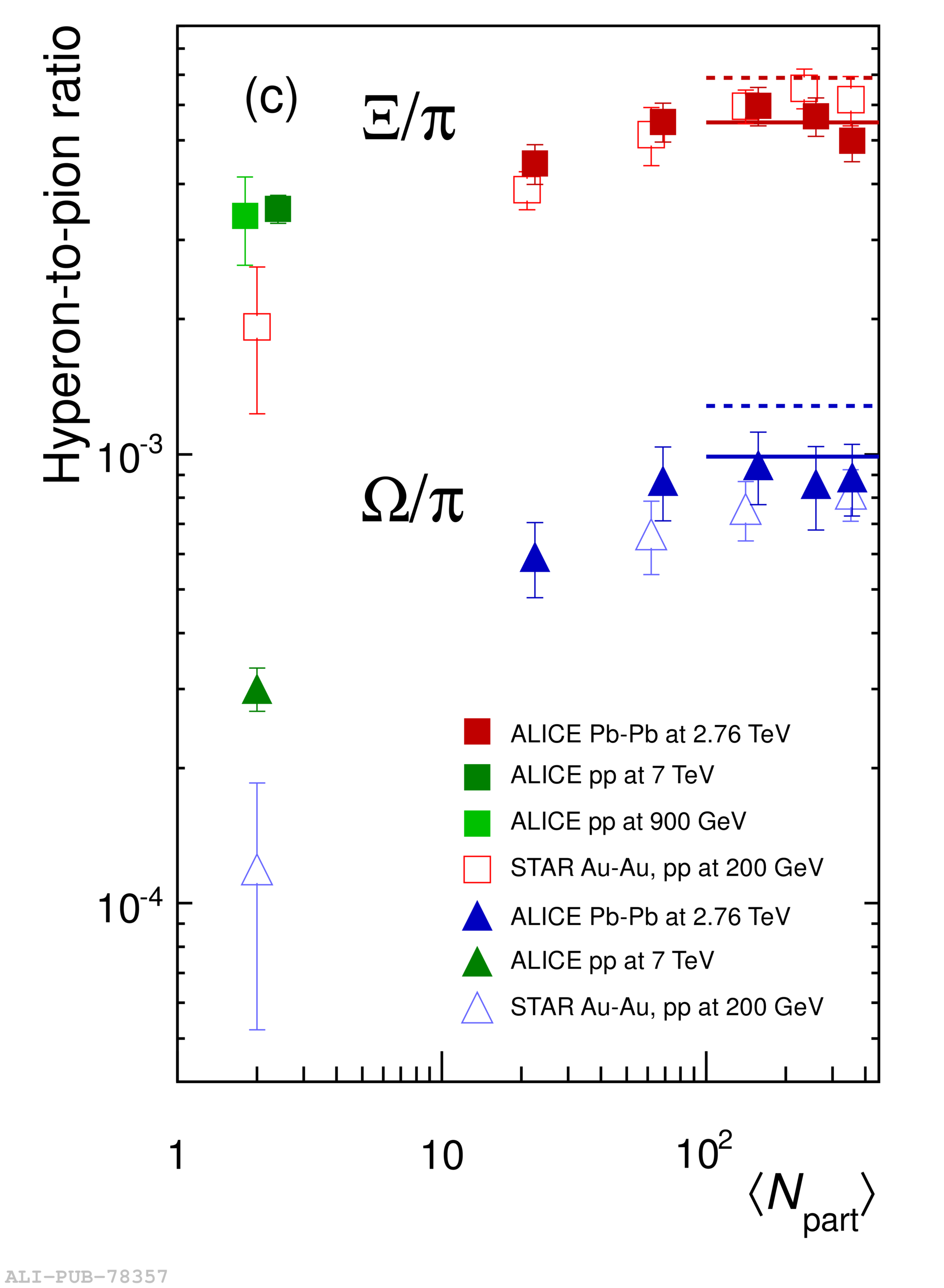}
\caption{(a), (b) Enhancements in $|y|$ $<$ 0.5 as a function of $\langle{N_{\rm part}\rangle}$ measured by ALICE and compared to SPS and RHIC data. The bars on the dotted line indicate the systematic uncertainties on the pp reference. (c) Hyperon-to-pion ratios as a function of $\langle{N_{\rm part}\rangle}$, for \mbox{A--A} and pp collisions at LHC and RHIC energies.}
\label{fig-1}  
\end{figure}

The strangeness enhancement is defined as the ratio between the yields in \mbox{Pb--Pb} collisions and those in \mbox{pp} interactions at the same energy, both normalized to the mean number of participants ($\langle{N_{\rm part}\rangle}$). The measured enhancement for multi-strange baryons is shown in Figure \ref{fig-1}a and \ref{fig-1}b as a function of $\langle{N_{\rm part}\rangle}$: it increases with centrality and with the strangeness content of the particle as already observed at lower energies, and decreases as the centre-of-mass energy increases, continuing the trend established between SPS and RHIC energies.

\begin{figure}
\centering
\includegraphics[width=15pc,clip]{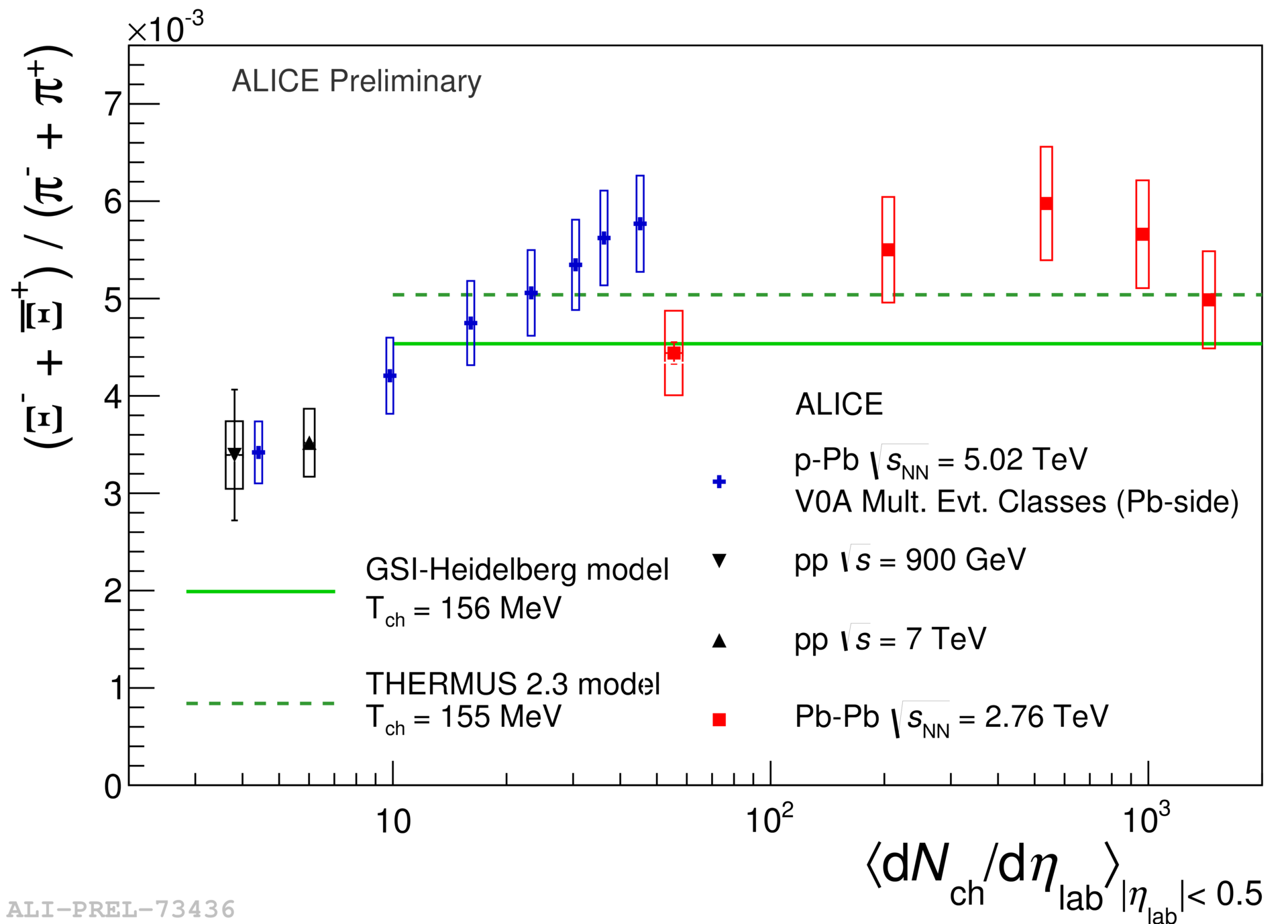}
\includegraphics[width=15pc,clip]{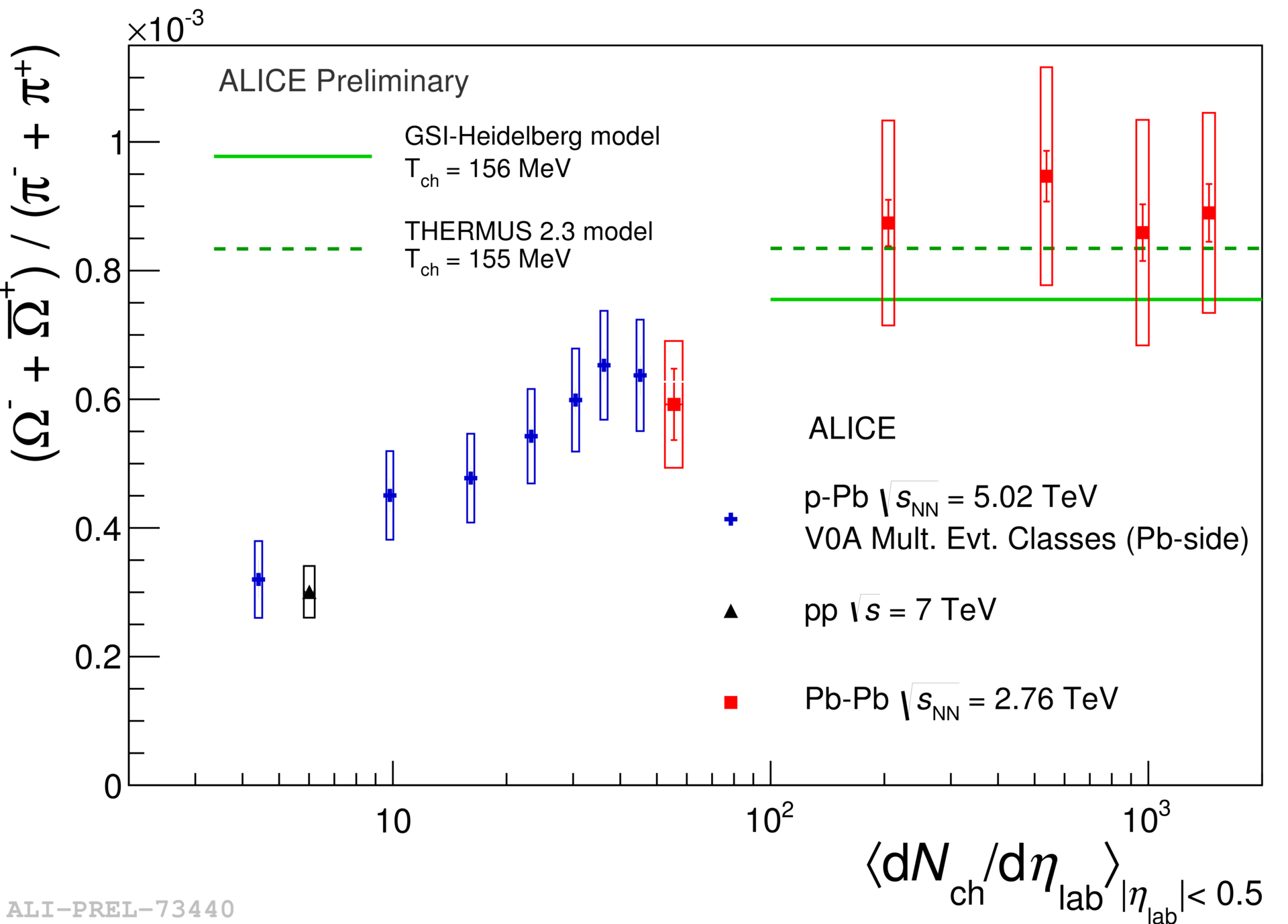}
\caption{$(\Xi^{-}+\overline{\Xi}^{+})/(\pi^{-}+\pi^{+})$ and $(\Omega^{-}+\overline{\Omega}^{+})/(\pi^{-}+\pi^{+})$ ratios as a function of charged-particle multiplicity density in \mbox{pp}, \mbox{p--Pb} and \mbox{Pb--Pb} collisions.}
\label{fig-2}  
\end{figure}

An alternative way to look at the strangeness enhancement, in order to factor out the general increase in multiplicity with \textit{N}$_{\rm part}$ from the actual increase in strange particle production is to study the hyperon-to-pion ratios, namely \mbox{$(\Xi^{-}+\overline{\Xi}^{+})/(\pi^{-}+\pi^{+})$} and \mbox{$(\Omega^{-}+\overline{\Omega}^{+})/(\pi^{-}+\pi^{+})$}, for \mbox{A--A} and pp collisions. These ratios are shown in Figure \ref{fig-1}c as a function of $\langle{N_{\rm part}\rangle}$, both at LHC and RHIC energies. The relative production of strangeness in pp collisions at the LHC is larger than at RHIC energy while it is almost the same at the two energies for \mbox{A--A} collisions. The decrease of enhancement with increasing energy is then caused by the change in \mbox{pp} collisions, in agreement with the hypothesis of progressive relaxing of canonical suppression in \mbox{pp} collisions with increasing energy. The hyperon-to-pion ratio increases when going from pp to \mbox{A--A}, showing a relative enhancement of strangeness production in \mbox{A--A} collisions (normalized to the pion yield) which is about half of that seen when normalizing to $\langle{N_{\rm part}\rangle}$. Figure \ref{fig-2} shows the recently measured \mbox{p--Pb} hyperon-to-pion ratios as a function of the charged-particle multiplicity density, compared with results from \mbox{pp} and \mbox{Pb--Pb} collisions.  An enhancement of the multi-strange baryons with multiplicity in \mbox{p--Pb} data is observed; it still increases with the strangeness content of the baryon, considering also the $\Lambda$/$\pi$ ratios published in \cite{lfpPb}.

In Figure \ref{fig-1}c the predictions for the hyperon-to-pion ratios from thermal models, based on a grand canonical approach are also shown: these predictions are provided by the GSI-Heidelberg model described in \cite{GSImodel} (full line, with a chemical freeze-out temperature parameter \mbox{T = 164 MeV}) and THERMUS described in \cite{THERMUS} (dashed line, with \mbox{T = 170 MeV}). The corresponding temperatures were obtained from extrapolations of RHIC results to LHC energies. Recently,  a global fit of all the hadron yields, measured by the ALICE Collaboration, has also been performed \cite{thermalRHICvsLHC} with the thermal grand canonical model; the chemical equilibrium temperature parameter obtained from the simultaneous fit is lower than the one extrapolated from low energy results and corresponds to about \mbox{T = 156 MeV}. The solid (GSI-Heidelberg model) and dashed (THERMUS) green lines in Figure 2 show ratios corresponding to chemical equilibrium temperatures of 156 MeV and 155 MeV, respectively. 

\begin{figure}[t!]
\centering
\includegraphics[width=16pc,clip]{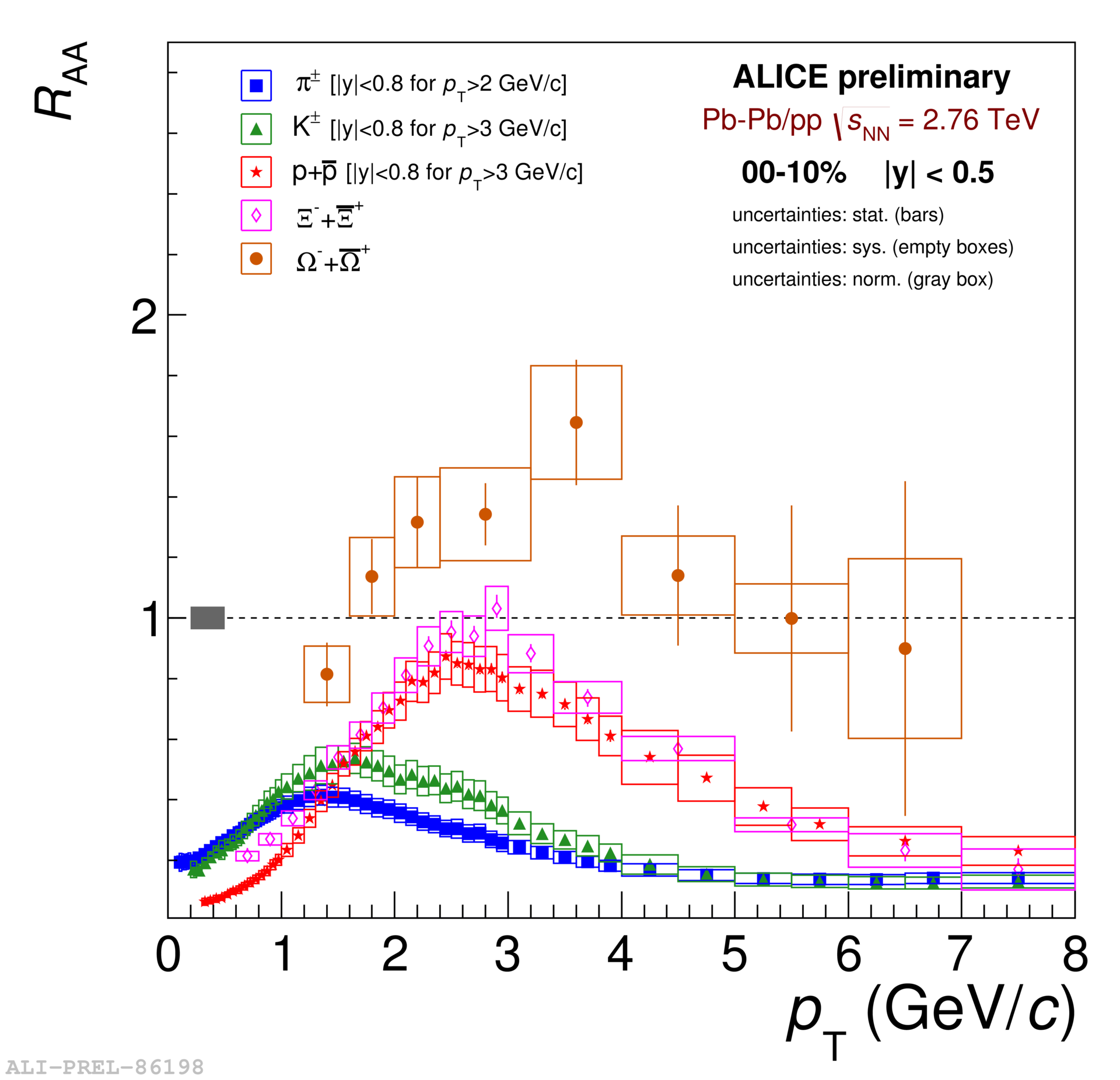}
\includegraphics[width=16pc,clip]{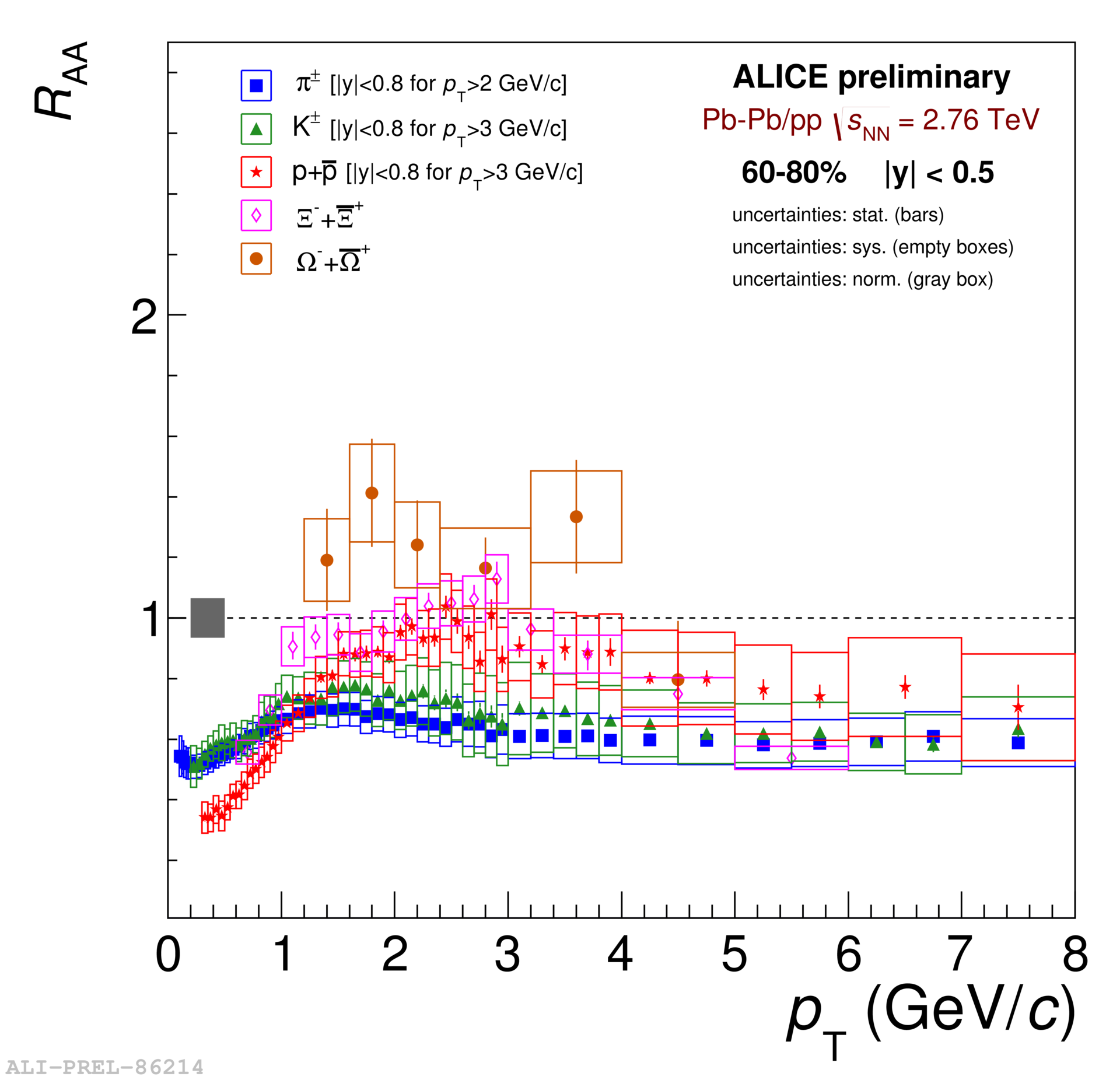} 
\caption{$\Xi$ and $\Omega$ nuclear modification factors compared to those for $\pi$, K and p in 0-10\% (left) and 60-80\% (right) centrality \mbox{$\sqrt{s_{NN}}$ = \mbox{2.76 TeV}} \mbox{Pb--Pb} collisions.}
\label{fig-3}  
\end{figure}

The nuclear modification factor ($R_{\rm AA}$) is defined as the ratio of the $p_{\rm T}$ spectra in \mbox{Pb--Pb}  and in \mbox{pp} collisions scaled by the number of nucleon-nucleon collisions \cite{centrality}. In Figure \ref{fig-3} the $R_{\rm AA}$ as a function of the $p_{\rm T}$ for multi-strange baryons, in the most central (0-10\%) and most peripheral (60-80\%) collisions, are shown and compared with those for lighter hadrons ($\pi$, K and p). In the most central class the reference spectra in \mbox{pp} collisions have been extended to the region \mbox{6 $<$ $p_{\rm T}$ $<$ 8 GeV/c} for the $\Xi$ and \mbox{5 $<$ $p_{\rm T}$ $<$ 7 GeV/c} for the $\Omega$, using a L\'evy-Tsallis fit function, to match the $p_{\rm T}$ reach of the spectra in \mbox{Pb--Pb} collisions. Looking at the most central events, the $R_{\rm AA}$ for $\Xi$ follows the same trend as the proton at high $p_{\rm T}$ (\mbox{$>$ 6 GeV/\textit{c}}), indicating that the suppression does not depend on the particle mass. At intermediate $p_{\rm T}$ there are indications of mass-ordering among the baryons, that can be interpreted as consequence of the radial flow \cite{Raa}. The $R_{\rm AA}$ for the $\Omega$ is above unity, which might be the result of the larger contribution of strangeness enhancement compared to  $\Xi$. When going from the most central to the most peripheral events the $R_{\rm AA}$ values and the relative differences between the particles are reduced. As expected, the strangeness enhancement effect for the $\Omega$ also becomes weaker in the most peripheral events.

\section{Conclusions}
\label{sec-4}
The production of the multi-strange baryons $\Xi^{-}$ and $\Omega^{-}$ and their anti-particles has been measured by the ALICE Collaboration in \mbox{pp}, \mbox{p--Pb} and \mbox{Pb--Pb} collisions. 
A strangeness enhancement has been observed in \mbox{Pb--Pb} collisions with similar features as in lower energy measurements. The decreasing trend with increasing centre-of-mass energy has been confirmed. The hyperon-to-pion ratios measured in \mbox{p--Pb} collisions bridge the values in \mbox{pp} to those in \mbox{Pb--Pb} and reveal a dependency of the strangeness enhancement on the charged-particle multiplicity. The comparison to the grand canonical thermal model expectations  for \mbox{T = 155 MeV} in \mbox{Pb--Pb} collisions yields, within a broad agreement with the measurements in \mbox{Pb--Pb} collisions, a tension between ratios for $\Xi$ and $\Omega$ in \mbox{p--Pb} collisions: while $\Omega/\pi$ only approaches the saturation level at the highest measured multiplicity, the $\Xi/\pi$ values at high multiplicity are comparable to the \mbox{Pb--Pb} values and lie above the equilibrium limits. Finally, the $R_{\rm AA}$ shows similar suppression for all the species at \mbox{$p_{\rm T} >$ 6 GeV/\textit{c}}. There is evidence of mass ordering at intermediate $p_{\rm T}$ and indication of a $R_{\rm AA}$ increase due to the larger strangeness enhancement for the $\Omega$.



\section*{References}
\begin{thebibliography}{13}

\bibitem{Rafelski} Rafelski J and M\"uller B 1982 {\it Phys. Rev. Lett.} {\bf 48} 1066 \\
Koch P, M\"uller B and Rafelski J 1986 {\it Phys. Rep.} {\bf 142} 167
\bibitem{strangEnhancNA57}
Antinori F {\it et al.} 2006 {\it J. Phys.} G {\bf 32} 427 \\ 
Antinori F {\it et al.} 2010 {\it J. Phys.} G {\bf 37} 045105 
\bibitem{strangEnhancSTAR}
Abelev B {\it et al.} 2008 {\it Phys. Rev.} C {\bf 77} 044908  
\bibitem{strangEnhancALICE}
Abelev B {\it et al.} 2014 {\it Phys. Lett. B} {\bf 728} 216  \\
Abelev B {\it et al.} 2014 {\it Phys. Lett. B} {\bf 734} 409  
\bibitem{RHICRaa}
Adcox K {\it et al.} 2002 {\it Phys. Rev. Lett.} {\bf 88} 022301  \\
Adler C {\it et al.} 2002 {\it Phys. Rev. Lett.} {\bf 89} 202301
\bibitem{ALICERaa}
Aamodt K {\it et al.} 2011 {\it Phys. Lett.} B {\bf 696} 30 
\bibitem{JINST}
Aamodt K {\it et al.} 2008 {\it JINST} {\bf 3} S08002
\bibitem{lfpPb}
Abelev B {\it et al.} 2014 {\it Phys. Lett.} B {\bf 728} 25-38
\bibitem{GSImodel}
Andronic A, Braun-Munzinger P and Stachel J 2009 {\it Phys. Lett.} B {\bf 673} 142 \\
Andronic A, Braun-Munzinger P and Stachel J 2009 {\it Phys. Lett.} B {\bf 678} 516 (Erratum).
\bibitem{THERMUS}
Cleymans J, Kraus I, Oeschler H, Redlich K and Wheaton S 2006 {\it Phys. Rev.} C {\bf 74} 034903
\bibitem{thermalRHICvsLHC}
Stachel J, Andronic A, Braun-Munzinger P and Redlich K 2014 {\it J. Phys. Conf. Ser.} {\bf 509} 012019
\bibitem{centrality}
Abelev B {\it et al.} 2013 {\it Phys. Rev.} C {\bf 88} 044909
\bibitem{Raa}
Abelev B {\it et al.}  2014 {\it Phys. Lett.} B {\bf 736} 196-207

\end{thebibliography}
\end{document}